\documentclass[twocolumn]{aastex701}
\usepackage{graphicx} 
\usepackage{amsmath}
\usepackage{amsfonts}
\usepackage{indentfirst}
\usepackage{multirow}
\usepackage{color}
\usepackage{diagbox}
\usepackage{hyperref}
\usepackage{natbib}
\usepackage{bbding}
\usepackage{float}

\begin{document}

\title{Impact of Subsurface Temperature Gradients on Emission Spectra of Airless Exoplanets: the Solid-state Greenhouse and Anti-Greenhouse}

\author{Xintong Lyu}
\affiliation{Peking University, Beijing, China}
\email[show]{XL: 1700011332@pku.edu.cn}

\author{Daniel D.B. Koll}
\affiliation{Peking University, Beijing, China}
\email[show]{DK: dkoll@pku.edu.cn}

\correspondingauthor{Daniel D.B. Koll}



\begin{abstract}
An emerging goal of exoplanet science is to constrain the surface composition of airless exoplanets. Without the protection of an atmosphere, these planets are likely covered by a powder-like regolith, similar to the Moon. Laboratory studies show that, under vacuum conditions, such regoliths can develop subsurface temperature gradients, also known as the solid-state greenhouse effect. This effect can significantly modify the emission features of airless bodies, but its potential impact on exoplanets is still unexplored. Here we derive analytic solutions of the two-stream radiative transfer equations with scattering, absorption, plus emission, and combine them with Mie theory calculations to model subsurface temperature gradients and emission spectra of airless exoplanets. The results show exo-regoliths can develop strong solid-state greenhouse or anti-greenhouse effects, with temperature gradients $>200$~K in the upper-most subsurface ($\mathcal{O}(100)\mu$m). These temperature gradients alter surface emission features, modify secondary eclipse depths by up to $\sim50\%$, and can produce higher-than-blackbody emission at some wavelengths. In addition, we study whether subsurface temperature gradients can be disentangled from other microscopic effects, such as changes in space weathering or particle size. At least in some cases, the co-existence of these effects makes it essentially impossible to distinguish different surface compositions within the precisions achievable by JWST. Overall, subsurface temperature gradients thus open potentially new ways to characterize surfaces of airless exoplanets, but they also complicate the interpretation of airless exoplanet spectra. In either case, their effect can be important and should be included in future modeling studies.
\end{abstract}

\section{Introduction}

With the launch of the JWST, astronomers are now able to characterize rocky exoplanets. Analyzing the secondary eclipse spectrum and phase curve of an exoplanet can reveal valuable insights into its surface or atmosphere \citep{seager2010a,hu2012theoretical,line2013systematic,line2014systematic}.

So far, JWST observations of rocky exoplanets have not definitively detected any atmospheres \citep{kreidberg2025first}. The most promising signs of atmospheres come from secondary eclipse observations of lava planets, which are exoplanets hot enough that their surfaces should be molten \citep{hu2024secondary,teske2025thick}. In contrast, most rocky exoplanets cool enough to have a solid surface appear compatible with airless bare rocks \citep[e.g.,][]{kreidberg2019absence,Zieba2023,xue2024jwst}.

By virtue of being airless, bare rock exoplanets allow us to directly study and characterize their surfaces. An emerging goal of exoplanet science is thus to use JWST observations to characterize the surface compositions and surface conditions of airless exoplanets, which trace the formation and evolution of rocky exoplanets over time \citep{hu2012theoretical,first2025,paragas2025}. Interpreting any observation, however, requires a model. Are current exoplanet surface models adequate for capturing the physical processes that are important on airless exoplanets?

Current bare-rock exoplanet models typically approximate exoplanetary surfaces as uniform, slab-like, layers, with empirical reflectivity and emissivity spectra determined from laboratory measurements \citep{hu2012theoretical,mansfield2019,hammond2024identifying,Lyu2024,coy2025population}. While this approach is useful in capturing the zero-th order spectral features specific to different geologic materials, it overlooks the complex conditions present on airless bodies.


Airless bodies in the Solar System are subject to intense stellar winds and frequent micrometeorite impacts, leading to rough, porous, and weathered regolith surfaces \citep{pieters2016space,henderson1997near}. When sunlight scatters off a regolith, it can lead to multiple phenomena including bidirectional reflectance and shadowing effects, which complicate the interpretation of remote spectra and phase curves \citep{hapke2012theory}. These effects are most important for observations at short wavelengths, and have subsequently been incorporated into some exoplanet surface models \citep{hu2012theoretical,Lyu2024,coy2025population}.

At longer wavelengths, where thermal emission dominates, regoliths also present surprisingly complex physics.  Laboratory experiments show that thermal emission spectra of powders depend on atmospheric pressure. Powder spectra measured at standard Earth-like pressure strongly differ from spectra measured in vacuum \citep{logan1970emission,logan1973compositional}. The reason is that powders allow starlight to penetrate and be absorbed at depth, whereas thermal emission only escapes near the powders's surface. In vacuum, a regolith thus generates and sustains its own subsurface temperature gradient \citep{henderson1994near}, a phenomenon that is also called the solid-state greenhouse \citep{matson1989solid}.

The importance of subsurface temperature gradients and of the solid-state greenhouse on airless bodies is well-appreciated in the Solar System literature \citep{henderson1994near,henderson1997near,kaufmann2007laboratory,wordsworth2019enabling}. In contrast, its potential impact on airless exoplanets has so far not been explored, raising a number of questions:
\begin{itemize}
    \item When do airless exoplanets develop a solid-state greenhouse, and are there any alternative states?
    \item How large is the impact of subsurface temperature gradients on JWST secondary spectra? 
    \item Are JWST observations sufficient to disentangle subsurface temperature gradients from other grain-size regolith effects?
\end{itemize}  
The goal of this paper is to explore how subsurface temperature gradients affect thermal observations of airless exoplanets and settle the above questions.



\section{Methods}
\label{sec: method}

To simulate the radiative flux of energy through a regolith, we use the two-stream radiative transfer equations \citep{henderson1997near,pierrehumbert2010principles,heng2017exoplanetary}. Collimated starlight impinges on the regolith at a cosine angle $\mu_0$, and then gets converted via scattering into diffuse radiation. The upward and downward diffuse radiances are governed by
\begin{subequations}
\begin{equation}
\begin{aligned}
    \overline{\mu}\frac{dI^+}{d\tau}
    =&I^{+}
    -\frac{a}{2}(1+g)I^{+}
    -\frac{a}{2}(1-g)I^{-}\\
    &-(1-g)\frac{a}{4\pi}F_s e^{-\tau/\mu_0}
    -(1-a)B[T(\tau)],
\end{aligned}
\end{equation}
\begin{equation}
\begin{aligned}
    -\overline{\mu}\frac{dI^-}{d\tau}
    =&I^{-}
    -\frac{a}{2}(1+g)I^{-}
    -\frac{a}{2}(1-g)I^{+}\\
    &-(1+g)\frac{a}{4\pi}F_s e^{-\tau/\mu_0}
    -(1-a)B[T(\tau)].
\end{aligned}
\end{equation}
\label{twostream eq}
\end{subequations}
In these equations, $I_{+}$ and $I_{-}$ represent the upward and downward diffuse radiance at a single wavelength. The parameter $\overline{\mu}$ is the average cosine angle of upward and downward radiation, which for diffuse radiation is $\overline{\mu}=1/2$ \citep{pierrehumbert2010principles}. The optical depth $\tau$ is
\begin{equation}
    \tau=z/d,
    \label{opdepth eq}
\end{equation}
where $z$ is depth and $d$ is the e-folding length, or skin depth, for radiation extinction (absorption and scattering) \citep{henderson1997near,henderson1994near}. Here, $a$ represents the single scattering albedo, $g$ the asymmetry factor, $F_s$ the stellar irradiance, $T$ is temperature, and $B(T)$ is the Planck function
\begin{equation}
    B(T)=\frac{2hc^2}{\lambda^5[\exp(hc/k_B\lambda T)-1]},
\end{equation}
where the physical constants $h$, $c$, $k_B$ have their standard meanings. In general, the two-stream parameters ($d,a,g,$...) all depend on wavelength $\lambda$.

In Equations~\ref{twostream eq}, the first term describes the attenuation of radiance with increasing optical depth. The second and third terms represent the forward and backward scattering of radiance in the upward and downward directions, respectively. The fourth term accounts for the conversion of collimated stellar radiance into diffuse radiance via scattering, while the final term describes thermal emission.

We found the following analytical solutions for the above two-stream equations, \( I_{+} \) and \( I_{-} \) (derived in Appendix \ref{appendix B}). The solutions are
\begin{subequations}
\label{eqn: solutions 1}
\begin{equation}
\begin{split}
    I^{+}=&\frac{1}{2}\bigg[
    \frac{1-a+\overline{\mu}\Gamma}{\overline{\mu}} \int_{\tau}^{\infty}B[T(\tau^\prime)]e^{-(\tau^\prime-\tau)\Gamma}d\tau^\prime\\
    &+(A_{-}-C_{-})e^{-\tau\Gamma}\\
    &+\frac{-1+a+\overline{\mu}\Gamma}{\overline{\mu}}
    \int_{0}^{\tau}B[T(\tau^\prime)]e^{-(\tau-\tau^\prime)\Gamma}d\tau^\prime\bigg]\\
    &+S_{+}e^{-\tau/\mu_0},
\end{split}
\end{equation}
\begin{equation}
\begin{split}
    I^{-}=&\frac{1}{2}\bigg[\frac{-1+a+\overline{\mu}\Gamma}{\overline{\mu}} \int_{\tau}^{\infty}B[T(\tau^\prime)]e^{-(\tau^\prime-\tau)\Gamma}d\tau^\prime\\
    &+(A_{-}+C_{-})e^{-\tau\Gamma}\\
    &+\frac{1-a+\overline{\mu}\Gamma}{\overline{\mu}}
    \int_{0}^{\tau}B[T(\tau^\prime)]e^{-(\tau-\tau^\prime)\Gamma}d\tau^\prime\bigg]\\
    &+S_{-}e^{-\tau/\mu_0},
\end{split}
\end{equation}
\end{subequations}

in which
\begin{subequations}
\label{eqn: solutions 2}
\begin{equation}
    \Gamma=\frac{\sqrt{(1-a)(1-ag)}}{\overline{\mu}},
\end{equation}
\begin{equation}
    S_{+}=\frac{(1-ag)+\frac{\overline{\mu}}{\mu_0}g-(1-a)g-\frac{\overline{\mu}}{\mu_0}}{\Gamma^2-\frac{1}{\mu_0^2}}\frac{aF_s}{4\pi\overline{\mu}^2},
\end{equation}
\begin{equation}
    S_{-}=\frac{(1-ag)+\frac{\overline{\mu}}{\mu_0}g+(1-a)g+\frac{\overline{\mu}}{\mu_0}}{\Gamma^2-\frac{1}{\mu_0^2}}\frac{aF_s}{4\pi\overline{\mu}^2},
\end{equation}
\begin{equation}
\begin{split}
    A_{-}=-\frac{(1-ag)}{(1-ag)+\overline{\mu}\Gamma}
    \bigg[\frac{-1+a+\overline{\mu}\Gamma}{\overline{\mu}} \\ \int_{0}^{\infty}B[T(\tau^\prime)]e^{-(\tau^\prime-\tau)\Gamma}d\tau^\prime+2S_{-}\bigg],
\end{split}
\end{equation}
\begin{equation}
\begin{split}
    C_{-}=-\frac{\overline{\mu}\Gamma}{(1-ag)+\overline{\mu}\Gamma}
    \bigg[\frac{-1+a+\overline{\mu}\Gamma}{\overline{\mu}} \\ \int_{0}^{\infty}B[T(\tau^\prime)]e^{-(\tau^\prime-\tau)\Gamma}d\tau^\prime+2S_{-}\bigg].
\end{split}
\end{equation}
\end{subequations}

Together, these formulas comprise a comprehensive model of radiative heat transfer in a planetary surface, including absorption, scattering, and thermal emission. In practice we evaluate these analytical solutions at individual wavelengths, then numerically integrate them over all wavelengths to arrive at the net upward and downward fluxes at each vertical layer. Then we use a numerical scheme to iterate the temperature profile as a function of depth until radiative equilibrium is reached. The emergent upward flux at the upper boundary yields the planet–star contrast ratio. We split the planet's dayside into seven concentric rings, each spanning $15^\circ$ in stellar zenith angle, and compute fluxes for each ring separately, to obtain the net secondary eclipse spectrum.
Note that solutions of the two-stream equations yield isotropic outgoing fluxes, so ignore bidirectional scattering and emission effects. This approximation is justified, as we focus on thermal emission spectra for which bidirectional effects are usually less important than for reflected light spectra \citep{hapke2012theory,hu2012theoretical,coy2025population}. In addition, we assume scattering is always isotropic ($g=0$).

To implement this model for different surface materials, we calculate the wavelength-dependent optical depth via Equation~\ref{opdepth eq} using Mie theory and refractive index data. Classic Mie theory is only valid for spherical particles that are well-separated, which is not necessarily true inside a regolith. Nevertheless, Mie theory has also been extended to simulate more complicated structures, such as fractal dust aggregates \citep{tazaki2016light}, and it has proven useful in modeling laboratory spectra of regolith-like powders \citep{lucey2011optical}. 
For comparison, recent work on airless exoplanets has largely been based on laboratory reflectance measurements. In this method one prepares regolith samples of a given composition and particle size, then measures the sample's reflectance spectrum, which is then inverted to infer an effective single-scattering albedo that can be used in surface models \citep{hu2012theoretical,paragas2025,first2025}. In contrast, our approach uses Mie theory to predict quantities such as single-scattering albedo from first principles. To obtain refractive index data we searched the rocks and minerals category of the Oxford Aerosol Refractive Index Database (\href{https://eodg.atm.ox.ac.uk/ARIA/data}{ARIA}) and chose three materials relevant for airless exoplanets: Basalt, Granite, and Hematite \citep{egan1975ultraviolet,toon1977physical}. Our default simulations assume all materials are geologically fresh (no space weathering), with a uniform particle diameter of 10$\mu$m, although we revisit these assumptions in Section \ref{Other Effects}.

Apart from radiative transfer, thermal conduction is also important in determining the subsurface's temperature profile. The conductive heat flux is
\begin{equation}
    F_c=-k \frac{dT}{dz},
    \label{conduction eq}
\end{equation}
where $k$ is the subsurface's thermal conductivity. Previous work explored the impact of background pressure on the emission spectra of regolith samples, so $k$ represented the net heat conductivity due to solid grains plus interstitial gas inside the regolith \citep{henderson1997near,kaufmann2007laboratory}. Here we focus on airless planets, so $k$ represents only the heat flow due to conduction between solid grains.
For simulations with non-zero $k$, we again use numerical iteration to find the temperature profile for which the net (radiative plus conductive) flux is constant with depth. Because $k$ on exoplanets is a priori unknown, we explore a wide range of thermal conductivities; for reference, a representative value for the lunar regolith is $k\sim10^{-4}$ to $10^{-2}~$W/(m K) \citep{pierrehumbert2010principles,yu2016thermal}.
In addition, both thermal conductivity $k$ and optical depth $\tau$ depend on the surface’s particle packing fraction \citep{henderson1997near}. Fewer particles along the light path reduce radiative interactions, and we adopt an occupancy rate of 0.3 to represent a partially porous regolith \citep{henderson1994near}.

The specific exoplanets we consider are TRAPPIST-1b and LHS 3844b. All simulations use Sphinx stellar spectra for TRAPPIST‑1 and LHS 3844 at 0.1$\mu$m resolution \citep{iyer2023sphinx}. The model's bottom boundary is set deep enough so radiative fluxes are constant with depth (typically, 1000$\mu$m). The assumed bottom boundary condition is that temperature gradients vanish, which implies zero heat flux from the planet's interior.

\section{Results}
\label{sec: results}

\subsection{Solid-state greenhouse versus anti-greenhouse}
\label{Vanilla Model}

\begin{figure*}[t]
\centering
    \includegraphics[width=\textwidth]{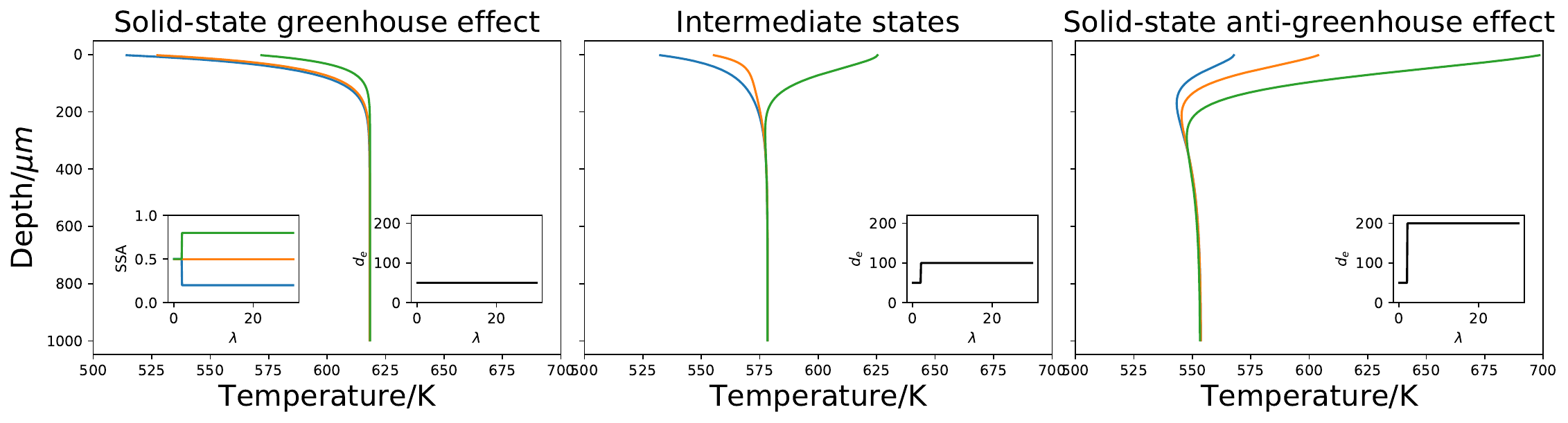}
    \caption{The solid-state greenhouse versus anti-greenhouse is primarily controlled by the relative extinction depth of radiation at short versus long wavelengths, $d_s\mu_0$ versus $d_l\overline{\mu}$. Insets show idealized semi-grey spectra of single-scattering albedo (SSA) and extinction depth ($d$). Left: A strong solid-state greenhouse develops for $d_s\mu_0> d_l\overline{\mu}$. Middle: Intermediate states with $d_s\mu_0 \approx d_l\overline{\mu}$. In this case temperature gradients are primarily governed by the wavelength-dependent single-scattering albedo. Right: A strong solid-state anti-greenhouse develops for $d_s\mu_0<d_l\overline{\mu}$. Here, subscripts `s' and `l' denote properties shortward and longward of 2$\mu$m. All profiles assume an instellation equal to that of TRAPPIST-1b's substellar point.
    }
    \label{dsdl}
\end{figure*}

Previous work on subsurface temperature gradients focused on the solid-state greenhouse effect \citep{henderson1994near,henderson1997near,hapke1996,kaufmann2007laboratory,wordsworth2019enabling}. The effect is analogous to the greenhouse effect in Earth's atmosphere, and arises when stellar radiation penetrates deeply into a medium (surface or atmosphere) before it is absorbed, whereas thermal emission to space is effectively absorbed by the medium and so can only escape to space from shallow layers. This configuration leads to warm temperatures at depth and cooler temperatures above.
However, there is no reason why a regolith might not exhibit a solid-state anti-greenhouse effect. A prime example of the anti-greenhouse effect is Titan's atmosphere, which absorbs stellar radiation more efficiently than thermal radiation \citep{mckay1991greenhouse,mckay1999analytic}, leading to a reversed temperature profile with warmer temperatures above and cooler temperatures below.

We therefore first consider when planetary surfaces exhibit a solid-state greenhouse versus anti-greenhouse. To keep things simple, we focus on pure radiative equilibrium and assume $a$ and $d$ are semi-grey, with different properties shortward versus longward of $2\mu$m (for a completely grey surface, the greenhouse versus anti-greenhouse are independent of $a$ and $g$ and only depend on $\mu_0$ versus $\overline{\mu}$; see Appendix~\ref{appendix C}). Inspection of Equations~\ref{twostream eq} further reveals that $d$ only occurs in combination with a zenith angle, so $d \mu$ controls the effective absorption depth of radiation at a given wavelength.
The influence of the single-scattering albedo $a$ on greenhouse versus anti-greenhouse effects is more complex, as $a$ influences both the absorption of heat and its distribution with depth. As $a$ increases, the surface scatters more and absorbs less. This enhanced scattering allows more radiation to be scattered deeper into the material, effectively increasing the depth $d$ at which energy is deposited. However, a larger $a$ also implies that a greater portion of incident energy is reflected back to space rather than absorbed, reducing the overall energy available for heating the subsurface. Using subscript `s' and `l' for shortwave and longwave wavelengths, we therefore expect that the greenhouse versus anti-greenhouse primarily depend on $a_s$, $a_l$, $d_s \mu_0$, and $d_l \overline{\mu}$.

Figure~\ref{dsdl} illustrates different temperature profiles in pure radiative equilibrium, calculated for the semi-grey $a$ and $d$ spectra shown in the small insets, and for a stellar spectrum and instellation equal to that of TRAPPIST-1b. In agreement with previous work on Titan's atmosphere \citep{mckay1991greenhouse,mckay1999analytic}, we find that the dominant parameter which governs the transition from greenhouse to anti-greenhouse effect is the relative penetration depth of shortwave versus longwave radiation, namely $d_s \mu_0$ versus $d_l \overline{\mu}$.

The first panel of Figure~\ref{dsdl} shows simulations in which shortwave radiation penetrates deeply into the regolith relative to longwave radiation, $d_s\mu_0 > d_l\overline{\mu}$. The relative difference is large enough that even at $a_l = 0.8$, the effective absorption depth of emission remains smaller than that of instellation. This results in a strong solid-state greenhouse effect, regardless of the exact value of $a$.

The second panel of Figure~\ref{dsdl} shows simulations with $d_s\mu_0 = d_l\overline{\mu}$. For equal shortwave and longwave penetration depths one might expect neither a strong greenhouse nor anti-greenhouse effect, so the subsurface should become isothermal \citep{mckay1991greenhouse}. However, this does not happen for two reasons. First, the semi-grey approximation is imperfect on hot rocky exoplanets like TRAPPIST-1b due to their high temperatures and red-shifted stellar spectra. Instellation and thermal emission thus partially overlap in wavelength space, weakening the distinction between $d_s\mu_0$ versus $d_l\overline{\mu}$. Second, once $d_s\mu_0\approx d_l\overline{\mu}$, greenhouse versus anti-greenhouse effects become controlled by the single-scattering albedo $a$. For $a_l<a_s$, the surface still exhibits a greenhouse effect. For $a_l=a_s$, the greenhouse effect becomes extremely weak (less than $20$K). Finally, for $a_l>a_s$, the temperature profile flips into an anti-greenhouse. This behavior is controlled by the effect of scattering, which in the last case increases the effective penetration depth of thermal emission above that of instellation.

The third panel of Figure~\ref{dsdl}  shows simulations with $d_l \overline{\mu} > d_s \mu_0$. In this limit, all simulations develop a strong anti-greenhouse in the upper-most surface. However, the temperature profiles reach a minimum around 200$\mu$m, and increase slightly again at greater depths. This phenomenon arises from the interplay of instellation, upward emission from deeper layers, and downward emission from upper layers. Imagine an isothermal temperature profile. The upward thermal flux, which originates from $z=\infty$, is constant with depth. In contrast, the downward thermal flux originates from the subsurface's upper layers and increases with depth. Meanwhile, the stellar flux is absorbed and so decreases with depth. This interplay between increasing downward thermal flux and decreasing instellation creates a minimum in absorbed flux, leading to a minimum temperature at intermediate depths.

Overall, whether a regolith develops a solid-state greenhouse or anti-greenhouse thus depends on its spectra of absorption depth $d$ as well as single-scattering albedo $a$. In the next section, we show that different surface materials match these expectations, with some materials (basalt, granite) exhibiting a strong greenhouse effect and other materials (hematite) exhibiting a strong anti-greenhouse effect.

\subsection{Simulations of TRAPPIST-1b and LHS 3844b}
\label{Thermal Conduction}

\begin{figure*}[t]
    \centering
    \includegraphics[width=\textwidth]{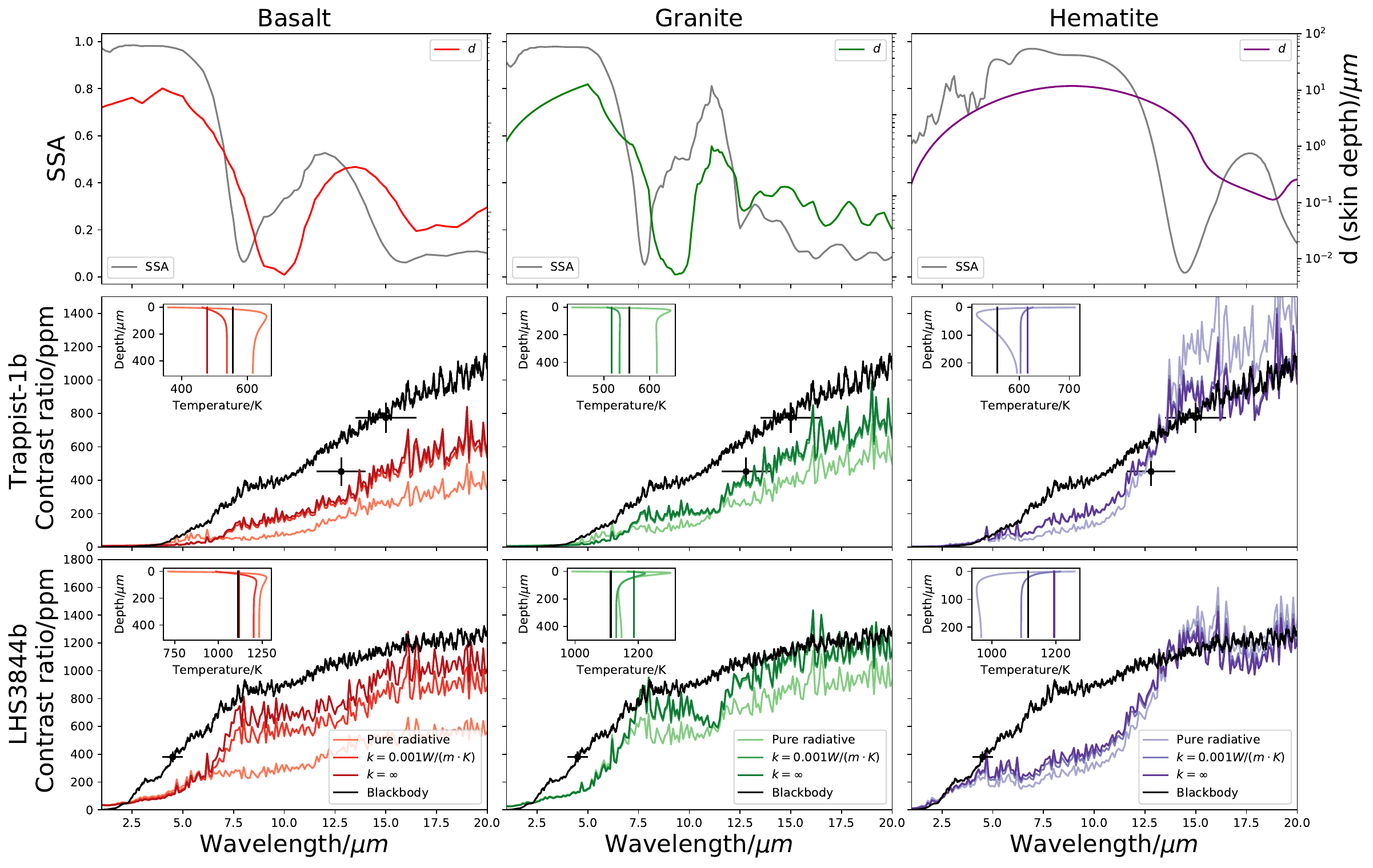}
    \caption{Variations in subsurface temperature gradients can modify the emission spectra of airless exoplanets by up to $\sim50\%$ at JWST wavelengths. Basalt and granite exhibit a strong solid-state greenhouse, whereas hematite exhibits a strong strong solid-state anti-greenhouse. Top row: For each material, spectra of skin depth (colored) and single-scattering albedo (grey). Middle and Bottom rows: Secondary eclipse spectra for TRAPPIST-1b and LHS 3844b. Insets shows the corresponding temperature profiles at the substellar point, for different thermal conductivity $k$. For reference, black points show Spitzer and JWST observations of TRAPPIST-1b and LHS 3844b.
    }
    \label{Compare}
\end{figure*}

We consider the impact of subsurface temperature gradients on TRAPPIST-1b and LHS 3844b for three different surface compositions: basalt, granite, and hematite. The simulated eclipse spectra are also compared against previous observations of TRAPPIST-1b \citep{ducrot2025combined} and LHS 3844b \citep{kreidberg2019absence}.
Simulations in this section now incorporate non-zero heat conduction via Equation~\ref{conduction eq}. Heat conduction always reduces temperature gradients compared to pure radiative equilibrium. Because $k$ on airless exoplanets is a priori unknown, we vary the thermal conductivity from $k$ from zero (pure radiative equilibrium), over a lunar-like $k=10^{-3}$ W/(m K), to infinity (isothermal surface).

Figure~\ref{Compare} shows the resulting temperature profiles at the substellar point and the dayside-integrated secondary eclipse spectra. We find that different materials exhibit strong greenhouse or anti-greenhouse effects, with temperature gradients of up to 200K in the upper-most subsurface ($\mathcal{O}(100)\mu$m). Near the surface, basalt and granite always develop a strong greenhouse effect. In contrast, hematite always shows a strong anti-greenhouse effect near the surface.

To understand why, consider the spectra of $d$ and $a$ shown in the top panel of Figure~\ref{Compare}.
%
These spectra should be compared to the relevant spectral ranges for absorption and emission. Both TRAPPIST-1 and LHS 3844 have effective temperatures around 3000K, leading to peak stellar absorption near 1$\mu$m \citep{gillon2017,vanderspek2019tess,zieba2022}. Meanwhile, TRAPPIST-1b has an equilibrium temperature of approximately 400K \citep{gillon2017}, so its thermal emission peaks around 7$\mu$m, whereas LHS 3844b has a higher equilibrium temperature of about 805K \citep{vanderspek2019tess}, so it emits most strongly close to 4$\mu$m.

Basalt has a larger skin depth and higher single-scattering albedo at short wavelengths, resulting in a clear solid-state greenhouse effect for both TRAPPIST-1b and LHS 3844b (see Fig.~\ref{dsdl}). Although basalt has a deeper skin depth at 4$\mu m$ than 1$\mu m$, the effective skin depth is smaller at 4$\mu m$ taking into account the angle of radiance. Granite exhibits a greenhouse effect for TRAPPIST-1b, and a mixed green-anti-greenhouse for LHS 3844b. Given its skin depth spectrum, we would expect a strong greenhouse effect for granite, and this is always the case at shallow depths. Meanwhile, granite's moderate anti-greenhouse effect for LHS 3844b is analogous to the non-monotonic temperature profiles in the right-most panel of Figure~\ref{dsdl}, which show temperature minima at intermediate depths. In the case of granite, this structure is flipped into a temperature maximum, leading to a moderate anti-greenhouse at depth.
Finally, hematite has a much smaller skin depth at short wavelengths than in the mid-infrared, consistent with its strong anti-greenhouse behavior in all simulations.

Next, we consider the impact of the thermal conductivity $k$. Thermal conduction can greatly reduce subsurface temperature gradients but, at least for lunar-like values of $k$, is insufficient to eliminate them. The inset panels in Figure~\ref{Compare} show that a larger $k$ always reduces vertical temperature contrasts. For a lunar-like thermal conductivity of $k =10^{-3}$ W/(m K), subsurface temperature gradients at depth are about half as big as they are in pure radiative equilibrium. In the upper-most subsurface, the impact of $k$ is more complicated. For granite, a lunar-like thermal conductivity is enough to greatly reduce the temperature difference in the top 100$\mu$m, making the surface's eclipse spectrum essentially identical to that of an isothermal surface (see below). In contrast, for basalt on LHS 3844b, a lunar-like $k$ results in an eclipse spectrum that is intermediate between those for $k=0$ and $k=\infty$.

Turning to the resulting eclipse spectra, we find that changes in thermal conductivity can have distinct and large spectral effects (see middle and bottom panels of Figure~\ref{Compare}). 
For basalt and granite, a larger thermal conductivity generally reduces secondary eclipse depths at wavelengths less than $5\mu$m, and increases them beyond $5\mu$m. Both materials absorb more effectively at longer wavelengths, so flux below $5\mu$m reflects thermal emission of the deep regolith ($\mathcal{O}(100)\mu$m), whereas flux beyond $5\mu$m reflects emission from the regolith's upper-most skin layer ($\mathcal{O}(10^{-1})\mu$m). 
This affects the broad spectral slopes of both materials, with broadband fluxes changing by up to a factor of 2 in response to large changes in thermal conductivity. Similarly, changes in conductivity also induce distinct spectral signatures. For example, for granite on TRAPPIST-1b, efficient thermal conduction flattens the eclipse spectrum between 7.5 and 11$\mu$m, while on LHS 3844b it introduces a new dip near 11$\mu$m.
Unlike basalt and granite, hematite emits much less at 4–13$\mu$m due to its high transparency and large single-scattering albedo. To satisfy energy conservation, hematite must emit more at long wavelengths. This leads to super-blackbody emission, in which hematite consistently emits more than a blackbody beyond 10$\mu$m, with the effect's magnitude modulated by the regolith's thermal conductivity -- a behavior distinct from that of basalt and granite. These spectral signatures, especially super-blackbody emission, may thus serve as detectable signs of subsurface temperature gradients on exoplanets.

Finally, comparing our simulations to existing Spitzer and JWST observations of TRAPPIST-1b and LHS 3844b, we find that current exoplanet observations are already precise enough to be sensitive to subsurface temperature effects. For example, a strong greenhouse effect on TRAPPIST-1b decreases its thermal emission at 15$\mu$m by up to 200ppm (about 1$\sigma$ with JWST) relative to an isothermal surface. For TRAPPIST-1b, we also find that hematite reproduces the observed slope between MIRI observations at 12.5$\mu$m and 15$\mu$m far better than basalt or granite. If so, a hematite-rich surface on TRAPPIST-1b should also induce super-blackbody emission beyond 15$\mu$m, which is a testable prediction for future observations. 


\subsection{Degeneracy between Subsurface Temperature Gradients, Space Weathering, and Particle Size Effects}
\label{Other Effects}

\begin{figure*}[t]
\centering
\includegraphics[width=\textwidth]{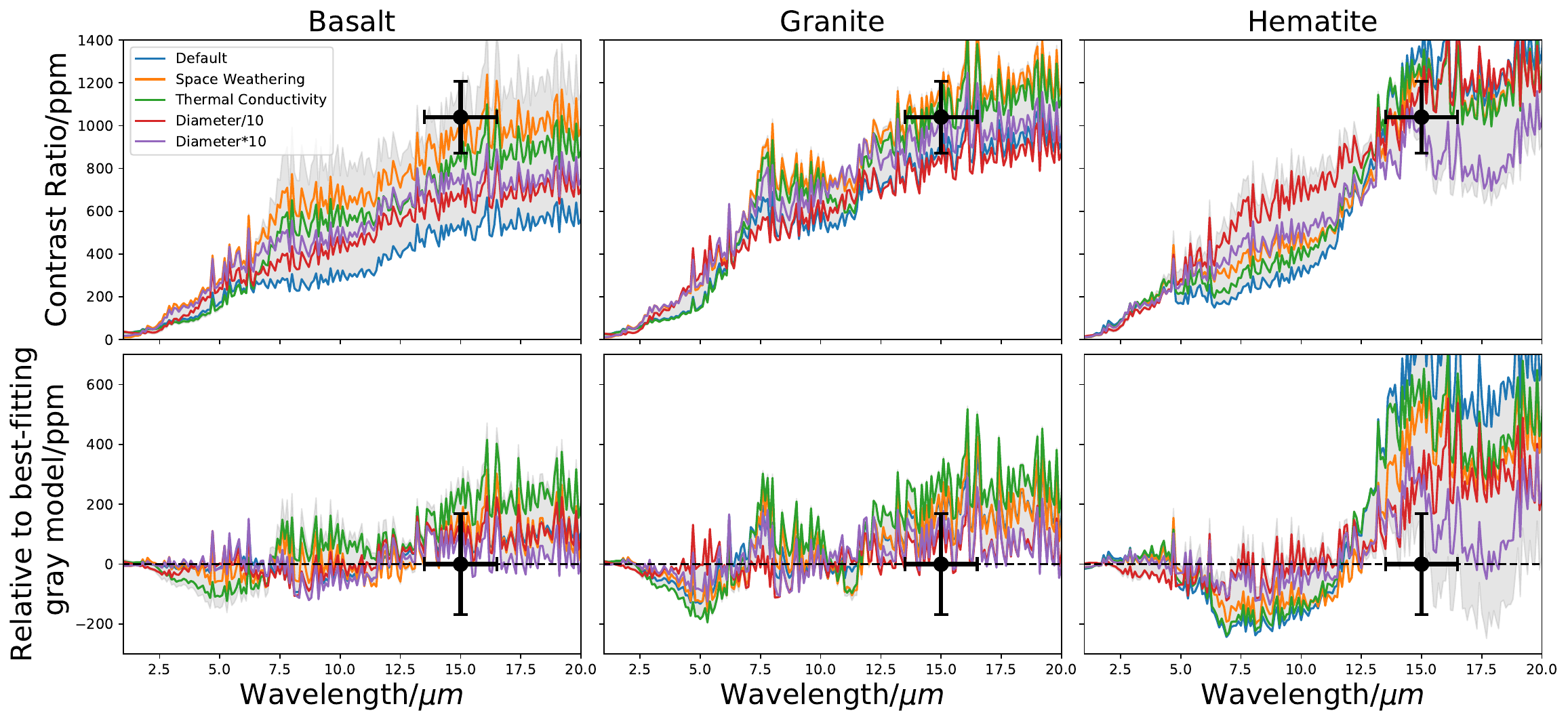}
\caption{Grain-scale variations via space weathering and particle size changes can modify secondary eclipse spectra as strongly as large changes in subsurface temperature gradients. Top rows: Simulations with variations in space weathering, thermal conductivity, or diameter changes compared to default simulations. Grey shows the possible range in eclipse spectra due to two co-varying effects (see Appendix~\ref{appendix A}).
Bottom rows: We take the spectra from top and subtract best-fit grey blackbodies, to isolate the induced spectral effect of different grain-scale variations. Black errorbars show a photon-noise estimate for possible 1$\sigma$ JWST precision at 15$\mu$m for one eclipse of LHS 3844b.}
\label{OVERALL}
\end{figure*}

Subsurface temperature gradients can have a significant influence on exoplanet spectra, but is it possible to distinguish them from other microscopic regolith processes that affect airless exoplanets?

Without the shielding of an atmosphere, exoplanet surfaces are likely modified by space weathering processes, such as exposure to solar wind, cosmic rays, and micro-meteoroid impacts \citep{hapke2001space,domingue2014,Lyu2024,coy2025population}.
Space weathering alters the optical properties of surface materials, typically darkening them and attenuating diagnostic absorption bands \citep{pieters2016space}. These effects arise from microscopic or nanoscopic changes in the regolith associated with the formation of metallic iron particles or introduction of graphite \citep{cassidy1975,pieters2000,syal2015darkening}.
Similarly, variations in particle size affect the scattering and absorption of radiation, thereby modifying a regolith's thermal emission characteristics \citep{moersch1995thermal,hunt1972variation}.

We therefore explore the potential degeneracy between subsurface temperature gradients and other microscopic regolith processes. Our default simulations assume fresh surface materials in radiative equilibrium ($k=0$), with a uniform particle diameter of 10$\mu$m. To explore the effects of space weathering, we add 5\% graphite as a space weathering agent. Compared to typical $\mathcal{O}(1)\%$ space weathering agent mixing ratios on Mercury and the Moon, this represents a strongly weathered surface \citep{fleischer1954,lucey1995,robinson2001,taylor2006}. To explore changes in particle diameter, we vary diameter between 1$\mu$m to 100$\mu$m, roughly encompassing the size distribution of lunar regolith \citep{colwell2007lunar,park2008characterization,rickman2012particle}. We then simulate overlapping variations in space weathering, thermal conductivity, and particle size, as detailed in Appendix \ref{appendix A}.

\begin{figure*}[t]
\centering
\includegraphics[width=\textwidth]{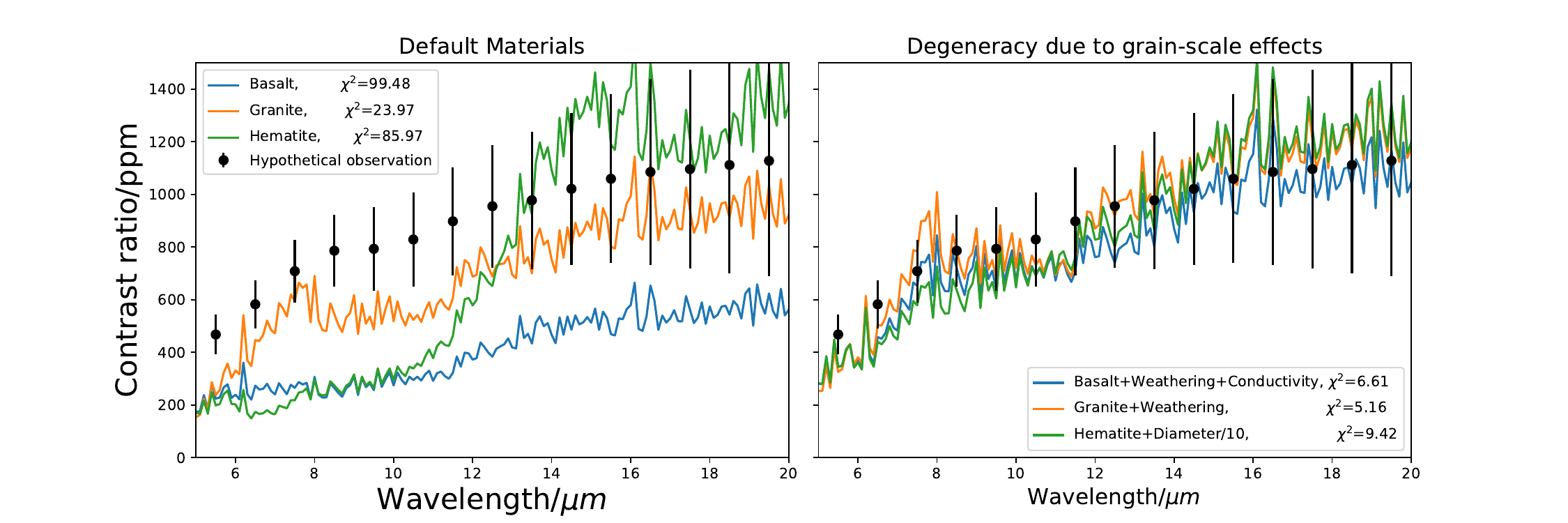}
\caption{Left: If grain-scale effects are held fixed, JWST can relatively easily distinguish between different surface compositions. Shown are secondary eclipse spectra of LHS 3844 for the three materials in Fig.~\ref{Compare} with default grain-scale conditions. Right: Allowing for co-variations in grain-scale effects can make it impossible to distinguish different surface compositions. Shown simulations were hand-picked to yield spectra that are effectively identical.
Black dots are hypothetical JWST observations, based on a grey body model with 0.1 albedo. Error bars show $1\sigma$ photon noise estimate for one eclipse of LHS 3844b with 1$\mu m$ bin size.
}
\label{degeneracy}
\end{figure*}

The results are shown in Figure~\ref{OVERALL}. Top panels illustrate the impact of single variations for each surface material (colored lines), as well as the maximum range of secondary eclipse depths when combining multiple variations (grey-shaded regions).

We find that variations in space weathering and particle size have a significant impact on exoplanet spectra, with either process potentially affecting the planet-star contrast ratio as much as variations in $k$. For example, at 15$\mu$m, changes in space weathering and particle size change the contrast ratio by more than 300 ppm for every surface material.
Part of this is not surprising. Stronger space weathering and larger particle sizes both act to darken surfaces \citep{pieters2016space,zaini2012effect,zhuang2023visible}, and thus reduce a planet's bond albedo. Energy balance then requires more thermal emission and a larger contrast ratio at long wavelengths. This bond albedo effect is most clearly visible for basalt: different microscopic variations cause the planet's emission to rise and fall, but the spectrum's shape remains relatively constant.

To take albedo changes into account, the bottom panels in Figure~\ref{OVERALL} subtract the best-fitting grey blackbody from each spectrum. We find that different microscopic changes not only modify a planet's bond albedo, they also strongly affect the shape and magnitude of spectral features. For example, the default hematite spectrum shows a strong spectral slope at 11-15$\mu$m, but this slope almost disappears for smaller particle sizes. Similarly, fresh basalt shows a spectral dip near 8$\mu$m, but this dip reverses into a bump once we add space weathering and change particle size.

Overlapping microscopic effects thus introduce significant ambiguity and can make it impossible to distinguish different surface compositions. Figure~\ref{degeneracy} shows model spectra of basalt, granite, and hematite for LHS 3844b, assuming default regolith parameters (left) versus a set with modified regolith parameters (right). As a proxy for hypothetical JWST observations, the black dots are based on a grey surface model with constant albedo of 0.1 while error bars show photon-noise for a single JWST eclipse, assuming perfect instrument efficiency. Photons are binned into 1$\mu$m bins and cover the wavelength range 5-20$\mu$m.

Considering only our default models, JWST is able to clearly characterize different surface compositions. For example, different surface materials exhibit flux differences of 400ppm at 8$\mu$m, and up to 1000ppm at 15$\mu$m, which are relatively easy to distinguish using JWST spectroscopy. However, once we allow for variations due to multiple microscopic processes, surface composition becomes almost indistinguishable. For example, space-weathered basalt with high thermal conductivity exhibits essentially the same emission spectrum as fresh hematite with a reduced particle size, and both spectra are virtually identical to a grey surface (see $\chi^2$ values in Fig.~\ref{degeneracy}, right).
Uncertainty in microscopic regolith processes can therefore significantly limit JWST's capacity to identify exoplanet surface composition solely from secondary eclipse measurements.

\section{Conclusions}
\label{sec: conclusion}

We combine new analytical solutions of the two-stream equations with Mie theory to model subsurface temperature gradients on airless exoplanets. Our results show that planetary regoliths can develop strong subsurface temperature gradients in the upper-most subsurface ($\mathcal{O}(100)\mu$m). Some regolith materials, such as basalt and granite, exhibit a strong solid-state greenhouse effect of up to several 100K. Other regolith materials, such as hematite, exhibit a similarly strong solid-state anti-greenhouse effect. The formation of a greenhouse versus anti-greenhouse primarily depends on the relative penetration depth of instellation versus thermal emission, $d_s\mu_0$ versus $d_l\overline{\mu}$, plus the regolith's wavelength-dependent single-scattering albedo $a$.

Our default simulations assume pure radiative equilibrium, which likely overestimates subsurface temperature gradients. In reality, grain-to-grain thermal conduction smooths out temperature gradients and drives the regolith towards a radiative–conductive equilibrium. 
However, unlike in the Solar System, there is no ground truth for the thermal conductivity $k$ of exoplanet regoliths. A lunar-like value for $k$ is insufficient to eliminate subsurface temperature gradients (Figure~\ref{Compare}). Moreover, $k$ might be even smaller on super-Earth exoplanets than on the Moon; grain-to-grain thermal conduction becomes less efficient for larger grains, and infrared observations of airless bodies in the Solar System suggests that larger bodies also have larger grain sizes \citep{gundlach2012outgassing,gundlach2013}. This calls for more high-temperature vacuum laboratory experiments to explore regolith physics in the parameter regime relevant for exoplanets \citep[e.g.,][]{paragas2025}. In the meantime, the most prudent approach is to treat microphysical regolith parameters such as $k$ as fundamentally unknown when interpreting exoplanet observations.

Turning to the effect of subsurface temperature gradients on exoplanet observations, we find that these gradients can modify secondary eclipse depths and spectral features by up to $50\%$. Importantly, for some materials such as hematite, subsurface temperature gradients can amplify the planet's infrared emission at specific wavelengths above that possible for a pure blackbody. TRAPPIST-1b is a potential candidate, as its current JWST spectrum is compatible with a hematite surface. If so, the planet might exhibit super-blackbody emission at wavelengths longer than $15\mu$m, which is a testable prediction. Overall, the signature of subsurface temperature gradients is thus potentially detectable by JWST, and could be used to help characterize the geological diversity of rocky exoplanets.

In other cases, however, the signature of subsurface temperature gradients cannot be distinguished from other microscopic regolith effects such as space weathering and particle size variations. We use the example of hypothetical LHS 3844b observations to show that, at least in some cases, these microscopic effects can make it virtually impossible to distinguish radically different surface compositions (basalt, granite, hematite) via JWST eclipses. The existence of subsurface temperature gradients thus also complicates the interpretation of exoplanet spectra. Breaking these degeneracies might require a variety of observational and modeling approaches. These include more obtaining even more precise eclipse data, combining secondary eclipses with JWST phase curves, reducing model uncertainties via laboratory experiments, or complementing JWST with other space- and ground-based observations.

\begin{acknowledgements}
This work was supported by NSFC grants 12473064 and 4211101266. We thank Laura Kreidberg and Elsa Ducrot for providing us with the Sphinx stellar spectra of LHS 3844 and TRAPPIST-1b used in previous papers. We thank Edwin Kite and Robin Wordsworth for helpful discussions about airless exoplanets and the solid-state greenhouse effect.
\end{acknowledgements}

\appendix

\section{Combinations of Simulations Considered In Figure~\ref{OVERALL}}
\label{appendix A}
\begin{table}[H]
\centering
\begin{tabular}{c||c|c|c|c}
Combinations & thermal conduction & space weathering & diameter*10 & diameter/10\\
\hline\hline
thermal conduction & \CheckmarkBold & \CheckmarkBold & \CheckmarkBold & \CheckmarkBold\\ \hline
space weathering & \CheckmarkBold & \CheckmarkBold & \CheckmarkBold & \CheckmarkBold\\ \hline
diameter*10 & \CheckmarkBold & \CheckmarkBold & \XSolidBold & \XSolidBold\\ \hline
diameter/10 & \CheckmarkBold & \CheckmarkBold & \XSolidBold & \XSolidBold\\ \hline
\end{tabular}
\end{table}

\section{Derivation of Analytical Solutions}
\label{appendix B}

Consider the two-stream equations for radiative transfer inside a regolith, 
\begin{subequations}
\begin{equation}
    \overline{\mu}\frac{dI^+}{d\tau}
    =I^{+}
    -\frac{a}{2}(1+g)I^{+}
    -\frac{a}{2}(1-g)I^{-}
    -(1-g)\frac{a}{4\pi}F_s e^{-\tau/\mu_0}
    -(1-a)B[T(\tau)]
\end{equation}
\begin{equation}
    -\overline{\mu}\frac{dI^-}{d\tau}
    =I^{-}
    -\frac{a}{2}(1+g)I^{-}
    -\frac{a}{2}(1-g)I^{+}
    -(1+g)\frac{a}{4\pi}F_s e^{-\tau/\mu_0}
    -(1-a)B[T(\tau)]
\end{equation}
\label{twostream}
\end{subequations}
in which $I_{+}$ and $I_{-}$ are the upward and downward diffuse radiance, $\overline{\mu}$ is the average cosine angle of diffuse radiation, and $\tau$ the optical depth, 
\begin{equation}
    \tau=z/d.
\end{equation}
Here $z$ is depth, $d$ the e-folding length or skin depth for radiation extinction (absorption and scattering), $a$ is the single scattering albedo, $g$ the asymmetry factor, $F_s$ the stellar irradiance, and $B(T)$ is the Planck function, 
\begin{equation}
    B(T)=\frac{2hc^2}{\lambda^5[\exp(hc/k_B\lambda T)-1]}.
\end{equation}

To simplify and solve these equations, we first define two variables:
\begin{subequations}
\begin{equation}
    2\phi=I^{+}+I^{-},
\end{equation}
\begin{equation}
    2\Delta \phi=I^{+}-I^{-},
\end{equation}
\end{subequations}
or equivalently
\begin{subequations}
\begin{equation}
    I^+=\phi+\Delta \phi,
\end{equation}
\begin{equation}
    I^-=\phi-\Delta \phi.
\end{equation}
\end{subequations}

By adding and subtracting Equations~\ref{twostream}, the equations become
\begin{subequations}
\begin{equation}
    \overline{\mu}\frac{d\phi}{d\tau}
    =(1-ag)\Delta\phi
    +g\frac{aF_s}{4\pi}e^{-\tau/\mu_0},
\end{equation}
\begin{equation}
    \overline{\mu}\frac{d\Delta\phi}{d\tau}
    =(1-a)\phi
    -\frac{aF_s}{4\pi}e^{-\tau/\mu_0}
    -(1-a)B(\tau).
\end{equation}
\end{subequations}

Taking another derivative, the coupled first-order differential equations become uncoupled second-order differential equations:
\begin{subequations}
\begin{equation}
    \frac{d^2\phi}{d\tau^2}
    =\Gamma^2\phi
    -\bigg[(1-ag)+\frac{\overline{\mu}}{\mu_0}\bigg]\frac{aF_s}{4\pi\overline{\mu}^2}e^{-\tau/\mu_0}
    -\Gamma^2B(\tau),
\end{equation}
\begin{equation}
    \frac{d^2\Delta\phi}{d\tau^2}
    =\Gamma^2\Delta\phi
    +\bigg[(1-a)g+\frac{\overline{\mu}}{\mu_0}\bigg]\frac{aF_s}{4\pi\overline{\mu}^2}e^{-\tau/\mu_0}
    -\frac{(1-a)}{\overline{\mu}}\frac{dB(\tau)}{d\tau},
\end{equation}
\end{subequations}
where $\Gamma$ is defined in Equation~\ref{eqn: solutions 2}.

The solutions are
\begin{subequations}
\begin{equation}
\begin{split}
    \phi=&\frac{1}{2}\bigg[\Gamma\int_{\tau}^{\infty}B[T(\tau^\prime)]e^{-(\tau^\prime-\tau)\Gamma}d\tau^\prime\\
    &+A_{-}e^{-\tau\Gamma}+
    \Gamma\int_{0}^{\tau}B[T(\tau^\prime)]e^{-(\tau-\tau^\prime)\Gamma}d\tau^\prime\bigg]\\
    &+\frac{(1-ag)+\frac{\overline{\mu}}{\mu_0}g}{\Gamma^2-\frac{1}{\mu_0^2}}\frac{aF_s}{4\pi\overline{\mu}^2}e^{-\tau/\mu_0},
\end{split}
\end{equation}
\begin{equation}
\begin{split}
    \Delta\phi=&\frac{1}{2}\bigg[\frac{(1-a)}{\overline{\mu}}\int_{\tau}^{\infty}B[T(\tau^\prime)]e^{-(\tau^\prime-\tau)\Gamma}d\tau^\prime\\
    &-C_{-}e^{-\tau\Gamma}
    -\frac{(1-a)}{\overline{\mu}}\int_{0}^{\tau}B[T(\tau^\prime)]e^{-(\tau-\tau^\prime)\Gamma}d\tau^\prime\bigg]\\
    &-\frac{(1-a)g+\frac{\overline{\mu}}{\mu_0}}{\Gamma^2-\frac{1}{\mu_0^2}}\frac{aF_s}{4\pi\overline{\mu}^2}e^{-\tau/\mu_0}.
\end{split}
\end{equation}
\end{subequations}

Substituting these solutions back into the original equations, parameters $A_{-}$ and $C_{-}$ satisfy:
\begin{equation}
    (1-ag)C_{-}=\overline{\mu}\Gamma A_{-}.
    \label{eqn: cminus aminus}
\end{equation}

Converting back to upward and downward radiances yields:
\begin{subequations}
\begin{equation}
\begin{split}
    I^{+}=&\frac{1}{2}\bigg[\frac{1-a+\overline{\mu}\Gamma}{\overline{\mu}} \int_{\tau}^{\infty}B[T(\tau^\prime)]e^{-(\tau^\prime-\tau)\Gamma}d\tau^\prime\\
    &+(A_{-}-C_{-})e^{-\tau\Gamma}
    +\frac{-1+a+\overline{\mu}\Gamma}{\overline{\mu}}
    \int_{0}^{\tau}B[T(\tau^\prime)]e^{-(\tau-\tau^\prime)\Gamma}d\tau^\prime\bigg]\\
    &+S_{+}e^{-\tau/\mu_0},
\end{split}
\end{equation}
\begin{equation}
\begin{split}
    I^{-}=&\frac{1}{2}\bigg[\frac{-1+a+\overline{\mu}\Gamma}{\overline{\mu}} \int_{\tau}^{\infty}B[T(\tau^\prime)]e^{-(\tau^\prime-\tau)\Gamma}d\tau^\prime\\
    &+(A_{-}+C_{-})e^{-\tau\Gamma}
    +\frac{1-a+\overline{\mu}\Gamma}{\overline{\mu}}
    \int_{0}^{\tau}B[T(\tau^\prime)]e^{-(\tau-\tau^\prime)\Gamma}d\tau^\prime\bigg]\\
    &+S_{-}e^{-\tau/\mu_0},
\end{split}
\end{equation}
\end{subequations}
where
\begin{subequations}
\begin{equation}
    S_{+}=\frac{(1-ag)+\frac{\overline{\mu}}{\mu_0}g-(1-a)g-\frac{\overline{\mu}}{\mu_0}}{\Gamma^2-\frac{1}{\mu_0^2}}\frac{aF_s}{4\pi\overline{\mu}^2},
\end{equation}
\begin{equation}
    S_{-}=\frac{(1-ag)+\frac{\overline{\mu}}{\mu_0}g+(1-a)g+\frac{\overline{\mu}}{\mu_0}}{\Gamma^2-\frac{1}{\mu_0^2}}\frac{aF_s}{4\pi\overline{\mu}^2}.
\end{equation}
\end{subequations}


At the top of the regolith, the diffuse downward radiance is zero, so that
\begin{equation}
    I^{-}(0)=\frac{1}{2}\bigg[\frac{-1+a+\overline{\mu}\Gamma}{\overline{\mu}} \int_{0}^{\infty}B[T(\tau^\prime)]e^{-(\tau^\prime-\tau)\Gamma}d\tau^\prime+
    (A_{-}+C_{-})\bigg]+S_{-}=0.
\end{equation}

When combined with Equation~\ref{eqn: cminus aminus}, this yields separate solutions for $A_{-}$ and $C_{-}$:
\begin{subequations}
\begin{equation}
    A_{-}=-\frac{(1-ag)}{(1-ag)+\overline{\mu}\Gamma}
    \bigg[\frac{-1+a+\overline{\mu}\Gamma}{\overline{\mu}} \int_{0}^{\infty}B[T(\tau^\prime)]e^{-(\tau^\prime-\tau)\Gamma}d\tau^\prime+2S_{-}\bigg],
\end{equation}
\begin{equation}
    C_{-}=-\frac{\overline{\mu}\Gamma}{(1-ag)+\overline{\mu}\Gamma}
    \bigg[\frac{-1+a+\overline{\mu}\Gamma}{\overline{\mu}} \int_{0}^{\infty}B[T(\tau^\prime)]e^{-(\tau^\prime-\tau)\Gamma}d\tau^\prime+2S_{-}\bigg].
\end{equation}
\end{subequations}

\section{Analytical Solutions for a Grey Medium in Radiative Equilibrium}
\label{appendix C}

Our analytical solutions in Appendix \ref{appendix B} apply at a single wavelength, and are used as such in the main text. However, starting from the two-stream equations, one can also derive solutions for a 'grey' medium (material properties $d$ and $a$ are independent of wavelength) in pure radiative equilibrium. The two-stream equations are
\begin{subequations}
\begin{equation}
    \overline{\mu}\frac{d\phi}{d\tau}
    =(1-ag)\Delta\phi
    +g\frac{aF_s}{4\pi}e^{-\tau/\mu_0},
\end{equation}
\begin{equation}
    \overline{\mu}\frac{d\Delta\phi}{d\tau}
    =(1-a)\phi
    -\frac{aF_s}{4\pi}e^{-\tau/\mu_0}
    -(1-a)B(\tau).
\end{equation}
\label{grey model two-stream}
\end{subequations}

To solve these equations, define two new parameters $F$ and $\Delta F$. $F$ is the sum of upward and downward diffuse fluxes, while $\Delta F$ is the net diffuse flux (difference between upward and downward fluxes):
\begin{subequations}
\begin{equation}
    F=\int_0^\infty 4\pi\overline{\mu}\phi \mathrm{d}\lambda,
\end{equation}
\begin{equation}
    \Delta F=\int_0^\infty 4\pi\overline{\mu}\Delta \phi \mathrm{d}\lambda .
\end{equation}   
\end{subequations}

Take these definitions back to Equations~\ref{grey model two-stream}, we can get
\begin{subequations}
\begin{equation}
    \frac{\mathrm{d}F}{\mathrm{d}\tau}=\frac{1-ag}{\overline{\mu}}\Delta F+ agF^se^{-\tau/\mu_0},
\end{equation}
\begin{equation}
    \frac{\mathrm{d}\Delta F}{\mathrm{d}\tau}=\frac{1-a}{\overline{\mu}}F-aF^se^{-\tau/\mu_0}-4(1-a)\sigma T^4,
\end{equation}
\end{subequations}
in which 

\begin{equation}
    F^s=\int_0^\infty F_s \mathrm{d}\lambda
\end{equation}
is the total instellation flux from central star.

Next, we take into account the collimated stellar flux,
\begin{subequations}
\begin{equation}
    F^\prime=F+\mu_0 F_s e^{-\tau/\mu_0},
\end{equation}
\begin{equation}
    \Delta F^\prime=\Delta F-\mu_0 F^s e^{-\tau/\mu_0}.
\end{equation}
\end{subequations}
so
\begin{subequations}
\begin{equation}
    \frac{\mathrm{d}F^\prime}{\mathrm{d}\tau}=
    \frac{1-ag}{\overline{\mu}}\Delta F^\prime 
    +(\frac{\mu_0}{\overline{\mu}}-1)(1-ag)F^se^{-\tau/\mu_0},
    \label{eqn: dFprime_dtau}
\end{equation}
\begin{equation}
    \frac{\mathrm{d}\Delta F^\prime}{\mathrm{d}\tau}=
    \frac{1-a}{\overline{\mu}}F^\prime
    +(1-\frac{\mu_0}{\overline{\mu}})(1-a)F^se^{-\tau/\mu_0}
    -4(1-a)\sigma T^4.
\end{equation}
\end{subequations}

Now $\Delta F^\prime$ is the net sum of all radiative fluxes (diffuse and collimated), defined positive upward. In radiative equilibrium this net sum of all fluxes has to be constant with depth, $\mathrm{d}\Delta F^\prime/\mathrm{d}\tau=0$, which leads to
\begin{equation}
    \frac{1-a}{\overline{\mu}}F^\prime
    +(1-\frac{\mu_0}{\overline{\mu}})(1-a)F^se^{-\tau/\mu_0}
    -4(1-a)\sigma T^4=0,
\end{equation}
which we solve for $\sigma T^4$:
\begin{equation}
    4\sigma T^4=\frac{1}{\overline{\mu}}F^\prime+(1-\frac{\mu_0}{\mu})F^s e^{-\tau/\mu_0}.
    \label{T-prof}
\end{equation}
This equation expresses the subsurface's temperature profile implicitly as a function of $F'$.

To arrive at an explicit equation for $\sigma T^4$, consider the regolith as a whole. Equilibrium requires that the surface receives as much flux as it emits, so the net flux has to be zero at the regolith's top boundary, $\Delta F^\prime(\tau=0)=0$. Combined with the requirement that the net flux remains constant with depth (see above), $\Delta F^\prime$ thus has to be 0 everywhere. Equation~\ref{eqn: dFprime_dtau} becomes
\begin{equation}
    \frac{\mathrm{d}F^\prime}{\mathrm{d}\tau}=
    (\frac{\mu_0}{\overline{\mu}}-1)(1-ag)F^se^{-\tau/\mu_0},
\end{equation}
which we integrate over $\tau$ to find,
\begin{equation}
    F^\prime=-\mu_0(\frac{\mu_0}{\overline{\mu}}-1)(1-ag)F^se^{-\tau/\mu_0}+C,
\end{equation}
where $C$ is a constant of integration.


Since $\Delta F^\prime(\tau=0)=0$ and $\mathrm{d}\Delta F^\prime/\mathrm{d}\tau=0$, $\Delta F^\prime$ is zero at any depth. At the surface, the incoming stellar flux induces instellation equals to radiation plus reflection. Besides, there is no downward fluxes in $F$ but only upward fluxes which equals to $F_s$ due to energy conservation. $F^\prime$, the sum of stellar and diffuse fluxes, is two times the instellation. 
\begin{equation}
    2F^s=-\mu_0(\frac{\mu_0}{\overline{\mu}}-1)(1-ag)F^s+C,
\end{equation}
\begin{equation}
    C=2F^s+\mu_0(\frac{\mu_0}{\overline{\mu}}-1)(1-ag)F^s.
\end{equation}

This allows us to solve for $F^\prime$, which is
\begin{equation}
    F^\prime=\mu_0(\frac{\mu_0}{\overline{\mu}}-1)(1-ag)F^s(1-e^{-\tau/\mu_0})+2F^s.
\end{equation}

Plugging back into Equation~\ref{T-prof}, we finally arrive at the temperature profile $\sigma T^4$ as a function of depth:
\begin{equation}
    4\sigma T^4=
    \frac{\mu_0}{\overline{\mu}}(\frac{\mu_0}{\overline{\mu}}-1)(1-ag)F^s(1-e^{-\tau/\mu_0})
    +\frac{2F^s}{\overline{\mu}}
    +(1-\frac{\mu_0}{\overline{\mu}})F^se^{-\tau/\mu_0}.
    \label{eqn: grey solution}
\end{equation}
Inspection of Equation~\ref{eqn: grey solution} shows that, for a perfectly grey surface, $T(\tau)$ can exhibit either a greenhouse or an anti-greenhouse effect, with the occurence of either only governed by the relative penetration depths of starlight versus thermal emission via $\mu_0$ versus $\overline{\mu}$.

\bibliography{zoterolibrary_latest,citation}
\bibliographystyle{aasjournalv7}
\end{document}